# A Nonparametric Method for Value Function Guided Subgroup Identification via Gradient Tree Boosting for Censored Survival Data

Pingye Zhang[1], Junshui Ma[1], Xinqun Chen[1] and Yue Shentu[1]

[1]MRL, Merck & Co., Inc., Rahway, NJ 07065, USA

Correspondence

Pingye Zhang, MRL, Merck & Co., Inc., Rahway, NJ 07065, USA.

Email: pingye.zhang@merck.com

## Abstract:

In randomized clinical trials with survival outcome, there has been an increasing interest in subgroup identification based on baseline genomic, proteomic markers or clinical characteristics. Some of the existing methods identify subgroups that benefit substantially from the experimental treatment by directly modeling outcomes or treatment effect. When the goal is to find an optimal treatment for a given patient rather than finding the right patient for a given treatment, methods under the individualized treatment regime framework estimate an individualized treatment rule that would lead to the best expected clinical outcome as measured by a value function. Connecting the concept of value function to subgroup identification, we propose a nonparametric method that searches for subgroup membership scores by maximizing a value function that directly reflects the subgroup-treatment interaction effect based on restricted mean survival time. A gradient tree boosting algorithm is proposed to search for the individual subgroup membership scores. We conduct simulation studies to evaluate the performance of the proposed method and an application to an AIDS clinical trial is performed for illustration.

# 1 Introduction

Although randomized clinical trials can establish the overall comparative treatment effect in a target patient population, patients may have heterogenous responses to treatments in reality. For example, a drug not benefiting the whole population may work for a subgroup. [1] An important goal of personalized medicine is to characterize treatment effect heterogeneity and to identify patient subgroups that exhibit differential treatment effects. The relevant statistical research can be commonly categorized into two frameworks: the first framework aims at identifying the right patient for a given treatment, often loosely grouped together as the subgroup identification methods; the second framework deals with identifying the right treatment for a given patient, [2] commonly referred as individualized treatment regime (ITR) in the literature.

For both randomized clinical trials and observational studies, an intuitive approach to identify patients who would have different clinical outcome under one treatment versus the other is to model the potential outcomes under either treatment. [3] Many methods followed this general concept to model the outcome conditional on patient characteristics and biomarkers. [4–14] Notably, machine learning methods have been proposed to achieve flexible and accurate prediction of clinical outcomes in this context (e.g. the Virtual Twin approach, [9] Counterfactual and Bivariate Random Forest, [13] and most recently a Gradient Boosting Tree-based approach[14] which is built on proportional hazard model for censored survival data). Since the outcomes can be affected by both main effects of baseline biomarkers and biomarker-treatment interactions in a very complex way, the performance of outcome modeling methods could be hurt by the model misspecification. In addition, the optimization criteria for predictive outcome modeling may not necessary align with the objective of identifying differential treatment effects. That is, what constitutes as the best predictive model for the clinical outcomes does not guarantee to optimize a reasonable value function for the subgroup identification or individual treatment recommendation problem.

An alternative strategy is to model the between-group treatment effect in the entire covariate space or in subsets of covariate space with enhanced treatment effect. [15–22] The treatment effect modeling methods bypass the estimation of main effects and is more robust to model misspecification, but similarly the final identified subgroups do not connect with an overall expected clinical benefit that arguably should be maximized.

Under the ITR framework, Qian and Murphy[23] introduced the notion of value function which reflects an expected clinical outcome for a treatment regime. The focus is to estimate the optimal treatment regime, if followed by all patients, would maximize the value function (assuming larger value represents better clinical benefit). Zhao et al.[24] ang Zhang et al.[25] transformed the value function optimization to a weighted classification problem known as outcome weighted learning. Numerous methods have been proposed under this framework, [26–38] with developments in the areas of robust estimation, variable selection, and sparse or interpretable treatment regime using tree-based methods. Among the clinical endpoints considered by ITR methods, the time-to-event or survival endpoints deserve some special attention. The ITR methods by construct utilize the individual survival outcome without the context of any specific candidate subgroups. Therefore, a censored survival outcome is either treated the same way as an observed event of the same duration in the outcome weighted learning approach, [31] or the explicit modeling of the censoring probability must be carried out in conjunction. [32] Neither approach is ideal. The first approach ignores the different information implied in a censored survival outcome, while the second approach involves additional modeling that is susceptible to model misspecification. In traditional survival analysis, censoring information under the assumption of noninformative censoring is interpreted in a population context, contributing to the number at risk at an event time point without any explicit modeling. Such interpretation extends naturally in the subgroup identification framework, where the censored survival data together with observed events define the survival distribution in a candidate subgroup.

While conventionally the survival data comparison in clinical trials are often based on log-rank test and cox proportional hazard model, restricted mean survival time (RMST) has received increasing attentions thanks to its intuitive interpretation, especially in the presence of non-proportional hazards. [39-41] For the subgroup identification problem in randomized clinical trials, the difference of RMST between two treatment groups under consideration has its unique appeal as the measure of treatment benefit. It is a model-free quantity that reflects the clinical intuition that wider space between two survival curves represents more survival benefit. In addition, the product of subgroup prevalence and the difference in RMST approximate a "total" measure of survival benefit from a public health perspective (the "mean" multiplied by the number of subjects). Furthermore, censoring information is accounted for naturally in the calculation of RMST.

In practice, it is of great interest to evaluate the effect of subgroup identification. Xu[31] proposed two quantities in which one measures the average experimental treatment effect in subgroup that experimental treatment leads to better clinical outcome (treatment performing subgroup) and the other quantity measures average control effect in subgroup that experimental treatment not leads to better outcome (treatment non-performing subgroup). An ideal subgroup identification would maximize both quantities. We extend Xu's idea by combining the two quantities together as a measure of the subgroup identification. Specifically, the new measure is defined as the experimental treatment effect in treatment performing subgroup weighted by the treatment performing subgroup prevalence minus the experimental treatment effect in treatment non-performing subgroup weighted by the treatment non-performing subgroup prevalence. The new quantity reflects a measure of subgroup-treatment interaction effect and an ideal subgroup identification would maximize it. For censored survival data, the experimental treatment effect can be estimated by the difference in RMST between the experimental treatment group and control group. A visualization of the proposed measure is presented in Figure 1.

In this paper, we propose a nonparametric method that connects a value function with the final subgroups identified. The value function is defined using the abovementioned measure of subgroup identification which reflects the subgroup-treatment interaction. The goal is to identify the treatment performing subgroup and its complementary subgroup, treatment non-performing subgroup, so that the differential treatment effects weighted by the prevalence of subgroups, measured by the value function, will be maximized. Instead of estimating the outcome or treatment contrast or utilizing outcome weighting at the patient-specific level, the proposed method compares difference in RMST in subgroups constructed by subgroup membership scores of patients. The subgroup membership scores are the parameters to be searched so that the treatment-subgroup interaction will be maximized. Gradient tree boosting is proposed to search for the optimal subgroup membership scores. Different from a typical use of gradient tree boosting solving a supervised classification problem with individual loss function (e.g. the Gradient Boosting Tree-based approach proposed by Sugasawa and Noma[14] uses negative log-Cox partial likelihood), the proposed value function does not involve individual label to evaluate misclassification error for individual patients. Our value function is based on measuring differential treatment effect at subpopulation level which may reduce variability coming from unpredictable sources of patient variability so that the overall prediction error could be reduced. Although individual loss function is not defined, the value function is differentiable with respect to individual subgroup membership score and the gradient of the value function can be used in gradient tree boosting.

The remainder of the paper is organized as follows. In section 2, we present the proposed method. We conduct simulation studies to evaluate performance of the proposed method in section 3. In section 4, we apply the proposed method to a randomized AIDS clinical trial. We conclude the paper with a discussion in section 5.

## 2 Materials and Methods

Assume we have a two-armed randomized controlled trial in which treatment is equally assigned to n patients from a population of interest. We let $A = 0$ or $1$ be the binary treatment indicator for control and experimental treatment, respectively. Furthermore, denote $T_1, \dots, T_n$ as the true survival times and $C_1, \dots, C_n$ the censoring times for the n patients. We define the observed survival time and censoring indicator as $X_i = \min(T_i, C_i)$, $\delta_i = I(T_i \leq C_i)$, and further assume $\mathbf{Z}$ as a q-dimensional baseline patient characteristic vector. The observed data are then $\{Y_i = (X_i, \delta_i), A_i, \mathbf{Z}_i\}$ for $i = 1,2, \dots, n$, which are assumed to be independently and identically distributed.

### 2.1 Proposed Value Function

We propose the below value function as a measure of the clinical interest associated with the subgroups to be identified.

$$V(\tau) = P[\tau(\mathbf{Z}) = 1]\{E[Y|A = 1, \tau(\mathbf{Z}) = 1] - E[Y|A = 0, \tau(\mathbf{Z}) = 1]\}$$

$$-P[\tau(Z) = 0]\{E[Y|A = 1, \tau(Z) = 0] - E[Y|A = 0, \tau(Z) = 0]\} \quad (1)$$

where $\tau$ is a rule to be estimated defining the subgroup membership according to $Z$, so that patients with $\tau(Z) = 1$ belong to the subgroup that experimental treatment leads to better clinical outcome (treatment performing subgroup), and patients with $\tau(Z) = 0$ belong to the subgroup that experimental treatment does not lead to better clinical outcome (treatment non-performing subgroup). The proposed value function can be viewed as a measure of differential experimental treatment effects across subgroups weighted by the subgroup prevalence. A visualization of the value function can be found in Figure 1.

To estimate $V(\tau)$ one could first to estimate $E[Y|A, \tau(Z)]$. For survival data, $E[Y|A, \tau(Z)]$ can be estimated using restricted mean survival time (RMST) estimate $\int_0^{t^*} S(t)\, dt$, [39–41] where $S(t)$ is the survival function estimate for the corresponding population and $t^*$ is a predefined cutoff time point for the RMST calculation. For example, $t^*$ can be set as the maximum observed follow-up time. For a given subgroup, $S(t)$ can be estimated by the Nelson–Aalen estimate. [42] For example, $S(t)$ for experimental treatment arm (A=1) in treatment performing subgroup can be estimated as below,

$$\hat{S}(t, A = 1, \tau = 1) = e^{-\Sigma_{t_k \leq t} \frac{\sum_{i=1}^n I(\tau_i=1)I(X_i=t_k)I(\delta_i=1)I(A_i=1)}{\sum_{i=1}^n I(\tau_i=1)I(X_i \geq t_k)I(A_i=1)}} \quad (2)$$

where $I(\tau_i = 1)$ is the indicator function determining whether the i-th patient belongs to the treatment performing subgroup or not. The problem remains is we do not have the observed data of subgroup membership beforehand. Here we define a subgroup membership score $p(Z_i)$ for each patient which can be viewed as an estimate of $P[\tau(Z_i) = 1]$: the probability of experimental treatment (A=1) is recommended to i-th patient given $Z_i$. $1 - p(Z_i)$ then is an estimate of $P[\tau(Z_i) = 0]$: the probability of control drug (A=0) is recommended to i-th patient given $\mathbf{Z}_i$. If we replace $I(\tau_i = 1)$ in (2) with $p(\mathbf{Z}_i)$ we can then estimate the survival function for experimental treatment arm in treatment performing subgroup as,

$$\tilde{S}(t, A = 1, \mathbf{p}) = e^{-\Sigma_{t_k \leq t} \frac{\sum_{i=1}^n p(\mathbf{Z}_i) I(X_i = t_k) I(\delta_i = 1) I(A_i = 1)}{\sum_{i=1}^n p(\mathbf{Z}_i) I(X_i \geq t_k) I(A_i = 1)}} \tag{3}$$

where $\mathbf{p} = [p_1(\mathbf{Z}) \quad \cdots \quad p_n(\mathbf{Z})]^T$. Similarly, survival functions for the remaining three subgroups defined by the agreement or disagreement of A and $\tau$ can be estimated using the formulas below

$$\begin{cases} \tilde{S}(t, A = 0, \mathbf{p}) = e^{-\Sigma_{t_k \leq t} \frac{\sum_{i=1}^n p(\mathbf{Z}_i) I(X_i = t_k) I(\delta_i = 1) I(A_i = 0)}{\sum_{i=1}^n p(\mathbf{Z}_i) I(X_i \geq t_k) I(A_i = 0)}} \\ \tilde{S}(t, A = 1, 1 - \mathbf{p}) = e^{-\Sigma_{t_k \leq t} \frac{\sum_{i=1}^n [1 - p(\mathbf{Z}_i)] I(X_i = t_k) I(\delta_i = 1) I(A_i = 1)}{[1 - p(\mathbf{Z}_i)] I \sum_{i=1}^n I(X_i \geq t_k) I(A_i = 1)}} \\ \tilde{S}(t, A = 0, 1 - \mathbf{p}) = e^{-\Sigma_{t_k \leq t} \frac{\sum_{i=1}^n [1 - p(\mathbf{Z}_i)] I(X_i = t_k) I(\delta_i = 1) I(A_i = 0)}{[1 - p(\mathbf{Z}_i)] I \sum_{i=1}^n I(X_i \geq t_k) I(A_i = 0)}} \end{cases}$$

where $\tilde{S}(t, A = 0, \mathbf{p})$ is the survival function estimate for control arm in treatment performing subgroup; $\tilde{S}(t, A = 1, 1 - \mathbf{p})$ is the survival function estimate for experimental treatment arm in treatment non-performing subgroup; $\tilde{S}(t, A = 0, 1 - \mathbf{p})$ is the survival function estimate for control arm in treatment non-performing subgroup. $E[Y|A, \tau(\mathbf{Z})]$ for the four subgroups can then be estimated as follows,

$$\begin{cases} \hat{E}[Y|A = 1, \tau(\mathbf{Z}) = 1] = \int_0^{t^*} \tilde{S}(t, A = 1, \mathbf{p}) \, dt \\ \hat{E}[Y|A = 0, \tau(\mathbf{Z}) = 1] = \int_0^{t^*} \tilde{S}(t, A = 0, \mathbf{p}) \, dt \\ \hat{E}[Y|A = 1, \tau(\mathbf{Z}) = 0] = \int_0^{t^*} \tilde{S}(t, A = 1, 1 - \mathbf{p}) \, dt \\ \hat{E}[Y|A = 0, \tau(\mathbf{Z}) = 0] = \int_0^{t^*} \tilde{S}(t, A = 0, 1 - \mathbf{p}) \, dt \end{cases}$$

With the above estimates for $E[Y|A, \tau(\mathbf{Z})]$, we propose to estimate $V(\tau)$ as below,

$$\hat{V} = \left(\sum_{i=1}^{n} p_i\right) \int_0^{t^*} [\tilde{S}(t, A=1, \boldsymbol{p}) - \tilde{S}(t, A=0, \boldsymbol{p})] \, dt - \left[\sum_{i=1}^{n}(1 - p_i)\right] \int_0^{t^*} [\tilde{S}(t, A=1, 1-\boldsymbol{p}) - \tilde{S}(t, A=0, 1-\boldsymbol{p})] \, dt \quad (4)$$

The goal is to estimate the subgroup membership score $p(\boldsymbol{Z}_i)$ for patients by maximizing the value function estimator in (4). Once $p(\boldsymbol{Z}_i)$ are estimated, we can stratify the overall population to treatment performing subgroup with patients with $p(\boldsymbol{Z}_i) > c$, and to treatment non-performing subgroup with patients with $p(\boldsymbol{Z}_i) \leq c$, where c is a cutoff, for example c can be 0.5.

## 2.2 Comparison Between Proposed Value Function and ITR Value Function

Note that under the setting of a two-armed 1:1 randomized controlled trial, (1) can be re-written as,

$$V(\tau) = E\left[\frac{YI(A = \tau(\boldsymbol{Z}))}{P(A|\boldsymbol{Z})}\right] - E\left[\frac{YI(A \neq \tau(\boldsymbol{Z}))}{P(A|\boldsymbol{Z})}\right] \quad (5)$$

To see that we first show the first term in (5) can be re-written as,

$$E\left[\frac{YI(A = \tau(\boldsymbol{Z}))}{P(A|\boldsymbol{Z})}\right] = \iiint_{YZA} YI[A = \tau(\boldsymbol{Z})] \, P(Y|A, \boldsymbol{Z}) P(\boldsymbol{Z})$$

$$= \int_Y YP[\tau = 1] P\{Y|A = 1, \tau(\boldsymbol{Z}) = 1\} + \int_Y YP[\tau = 0] P\{Y|A = 0, \tau(\boldsymbol{Z}) = 0\}$$

$$= P[\tau = 1] E[Y|A = 1, \tau(\boldsymbol{Z}) = 1] + P[\tau = 0] E[Y|A = 0, \tau(\boldsymbol{Z}) = 0] \quad (6)$$

Similarly, the second term in (5) can be re-written as,

$$E\left[\frac{YI(A \neq \tau(\boldsymbol{Z}))}{P(A|\boldsymbol{Z})}\right] = P[\tau = 1] E[Y|A = 0, \tau(\boldsymbol{Z}) = 1] + P[\tau = 0] E[Y|A = 1, \tau(\boldsymbol{Z}) = 0] \quad (7)$$

Therefore, subtracting (7) from (6) one can get the proposed value function in (1). Note that the first term in (5) is the widely used value function proposed by Qian and Murphy[23] under the ITR framework and is maximized when

$$\tau(\mathbf{Z}) = \begin{cases} 1, E(Y|A = 1, \mathbf{Z}) > E(Y|A = 0, \mathbf{Z}) \\ 0, E(Y|A = 1, \mathbf{Z}) \leq E(Y|A = 0, \mathbf{Z}) \end{cases} \quad (8)$$

Finding $\tau$ maximizing (6) is equivalent to finding $\tau$ minimizing (7) so the proposed value function in (1) will also be maximized when $\tau$ follows the equation (8). But different from comparing the predicted value of $E(Y|A = 1, \mathbf{Z})$ versus $E(Y|A = 0, \mathbf{Z})$ or directly maximizing the value function utilizing individual level outcome weighting, we propose to take apart the value function in (1) into expected outcome at subgroup level and directly maximize the value function comparing mean survival time among subpopulations defined by the agreement or disagreement of A and $\tau$. Instead of estimating treatment effect at individual level, we focus on treatment effect at subpopulation level to search for the optimizer for $\tau$.

## 2.3 Estimation via Gradient Tree Boosting

The gradient tree boosting is an ensemble method that constructs a predictive model by additive expansions of decision trees. [43] For the proposed method, we define $F_i$ as a logit function of $p(\mathbf{Z}_i)$ for patient $i = 1, \dots, n$.

$$F_i = \log\left[\frac{p(\mathbf{Z}_i)}{1 - p(\mathbf{Z}_i)}\right] \quad (9)$$

The final prediction $\hat{F}_i$ of $F_i$ follows an additive expansion of K base tree functions.

$$\hat{F}_i = \sum_{k=1}^{K} f_k(Y_i, A_i, \mathbf{Z}_i) \quad (10)$$

where $f_k$ is the k-th base tree function and K is the number of trees. To learn the set of base tree functions, the following regularized objective function is minimized at each of the K iterations. [44]

$$obj^{(k)} = \mathcal{L}\left(\left[\hat{F}_1^{(k-1)} + f_k(Y_1, A_1, \mathbf{Z}_1), \cdots, \hat{F}_n^{(k-1)} + f_k(Y_n, A_n, \mathbf{Z}_n)\right]^\mathrm{T}\right) + \Omega \quad (11)$$

where $\mathcal{L}$ is a differentiable loss function we want to minimize, $\Omega$ is a penalty function that penalizes the complexity of the tree functions and $\hat{F}_i^{(k-1)} = \sum_{j=1}^{k-1} f_j(Y_i, A_i, \mathbf{Z}_i)$ is the prediction up to the (k-1)-th base tree function. We present the proposed value function estimator in (4), here we simply set the loss function $\mathcal{L} = -\hat{V}$, the negative value function. We use the same penalty function as defined in xgboost. [44] Note that we do not define an individual loss function, so our loss function is not a direct summation of individual loss measuring counterfactual treatment effect at individual level. Instead, the loss function $\mathcal{L}$ is a measure of differential treatment effect at subpopulation level. The loss function $\mathcal{L}$ is differentiable with respect to $F_i$ and the first-order approximation can be used to optimize the objective function. Specifically, at the k-th iteration, the first-order gradient of the loss function evaluated at $\hat{F}_i^{(k-1)}$, $g_i^k = (\partial \mathcal{L}/\partial F_i)_{F_i=\hat{F}_i^{(k-1)}}$, can be used as the individual target label in the k-th tree to construct the prediction of $f_k(Y_i, A_i, \mathbf{Z}_i)$. We provide the derivation of the first-order gradient in the supplementary materials. The goal is to optimize the final prediction for $F_i$ to minimize the objective function and we can classify patients based on their estimated subgroup membership score $p(\mathbf{Z}_i)$. The numerical optimization can be implemented with commonly used software such as "xgboost". [44]

## 3 Simulation Studies

We conduct simulation studies to evaluate the performance of the proposed method. We consider two settings: in the first setting we assume no prognostic variables with marginal effects are presented; in the second setting we assume that 4 prognostic variables are involved in the underlying survival model. Six scenarios of subgroup patterns are considered under both settings.

Specifically, we consider a two-armed 1:1 randomized clinical trial with a uniform enrollment of 12 months followed by an 18 months study follow-up. We assume experimental treatment effects differ between two subgroups: treatment performing subgroup (Treatment > Control) and the treatment non-performing subgroup (Treatment ≤ Control). For each scenario we generate n=500 independent survival time samples based on the following models, respectively:

Scenario 1: $T = e^{\beta_0 + AS_1 + Z\beta_z + \sigma_0 \varepsilon}$

Scenario 2: $T = e^{\beta_0 + A(S_1 - S_2) + Z\beta_z + \sigma_0 \varepsilon}$

Scenario 3: $T = e^{\beta_0 + 2A\{I[-0.67 \leq S_1 < 0.67](e^{-S_1^2} - 0.4) + I[S_1 < -0.67 \cup S_1 \geq 0.67](e^{-S_1^2} - 0.8)\} + Z\beta_z + \sigma_0 \varepsilon}$

Scenario 4: $T = e^{\beta_0 + A\{2I[(-1.07 \leq S_1 < 1.07) \cap (-1.07 \leq S_2 < 1.07)] - 1\} - (Z\beta_z)^2 + \sigma_0 \varepsilon}$

Scenario 5: $T = e^{\beta_0 + A\{2I[(S_1 \geq 0.67) \cup (-0.67 \leq S_1 < 0)] - 1\} - Z^2 \beta_z + \sigma_0 \varepsilon}$

Scenario 6: $T = e^{\beta_0 + A\{2I[(S_1 \geq 0 \cap S_2 \geq -0.67) \cup (S_1 < 0 \cap S_2 < -0.67)] - 1\} - Z^2 \beta_z + \sigma_0 \varepsilon}$

where $S_1$ and $S_2$ are predictive variables involved in the determination of the above two subgroups and are generated along with the covariates $\mathbf{Z} = (z_1, \ldots, z_q)$ using a mean zero multivariate normal distribution with a compound symmetric variance-covariance matrix $\mathbf{\Sigma} = (1 - \rho)\mathbf{I}_{(q+2)} + \rho \mathbf{1}'\mathbf{1}$, and $\varepsilon \sim N(0,1)$. We let $q = 50, \rho = \frac{1}{3}, \beta_0 = \sqrt{6}, \sigma_0 = 0.4, \beta_Z = [\beta_1 \cdots \beta_q]^T = [0 \cdots 0]^T$ in setting 1 and $\beta_Z = [0.4, 0.4, 0.4, 0.4, 0 \cdots 0]^T$ in setting 2. The treatment indicator A is generated as 0 or 1 with equal probability at random. Exponential censoring is simulated assuming a yearly dropout rate of 10%. Patients who do not have events are censored at the end of study. The censoring rate is around 30% in setting 1 and is around 35% in setting 2. The six scenarios are summarized in Figure 2. In scenario 1, greater value in $S_1$ leads to greater experimental treatment benefit when $S_1 > 0$, and experimental treatment turns harmful when $S_1 < 0$ and the magnitude of treatment harmfulness increases as value of $S_1$ decreases. The two subgroups can be perfectly separated by the linear boundary $S_1 = 0$. In scenario 2, the linear boundary $S_1 = S_2$ can perfectly separate the two subgroups. The size of experimental treatment benefit or harmfulness changes along both the $S_1$ axis and $S_2$ axis. In scenario 3, the two subgroups can be separated by a nonlinear "U" shaped boundary. The magnitude of experimental treatment benefit increases when $S_1$ gets closer to 0 and the magnitude of experimental treatment harmfulness increases when $S_1$ gets farther away from 0.67 or -0.67. In scenario 4, Treatment > Control subgroup is "enclaved" by the Treatment ≤ Control subgroup. In scenario 5, the two subgroups can be separated by a "S" shaped nonlinear boundary. In scenario 6, the two subgroups are separated by two "L" shaped boundaries. In all scenarios, the Treatment > Control subgroup is simulated to be 50% of the total population.

We compare the proposed method with four existing methods. Tian et al.[20] proposed a regression-based method with modified covariates. Lasso-regularized cox regression is implemented using the R package "glmnet".[45] Fivefold cross-validation is used to find the optimal $l_1$ penalty parameter. This method is under the class of treatment effect modeling. Under the class of outcome modeling, Foster et al. proposed a 2-step approach with step-1 estimating the conditional outcome and step-2 fitting classification tree to identify subgroups. Under their framework, we first train a random survival forests model[46] with treatment by covariate interaction terms included as input covariates. We then use predicted survival probabilities to compute the difference in RMST estimates between the two arms for each of the patient in training set. At step-2, we train a classification random forest model[47] as the final prediction model with the dichotomized difference in RMST estimates obtained from step-1 as the target label. Both random survival forests and classification random forests are built with 1000 trees using the R package "randomForestSRC".[46] We refer to this method as "VT". In addition,

Sugasawa and Noma[14] proposed to estimate the conditional outcome via gradient boosting trees and compute the individual treatment effect as the difference of survival probabilities between two arms for some fixed time $t_0$. We choose $t_0$ to be 20 months which is about the 75$^{th}$ percentile of the simulated censored survival times. Following what they did in their paper, we fit gradient boosting trees separately to the experimental treatment and control groups based on the partial likelihood function and include $q$ covariates and their interactions in the models. We set the maximum number of trees to 2000 and the optimal number of trees is selected via five-fold cross-validation. We classify a patient into treatment performing subgroup if the predicted individual treatment effect is greater than zero. We refer to this method as 'SGBT' and use R package 'GBM' to implement it. We also include an outcome weighted learning method, regularized outcome weighted subgroup identification (ROWSi), proposed by Xu et al.[31] under the framework of outcome weighted learning. We use the 1-step version of their method without implementing the pre-screening group Lasso procedure as all variables are continuous in our simulation. R package "glmnet"[45] is used to implement the method. We implement our proposed method using the R package xgboost[44] which allows user defined loss function and evaluation function. We set the second order gradient to be 0.001 to implement first order gradient boosting. Note that since we set the second order gradient to be 0.001 we set the "min_child_weight"=0 to allow sufficient tree partitions. Five-fold cross-validation is used to find the optimal values for the step size shrinkage, maximum depth of a tree and number of trees. The other xgboost parameters are kept at default values. In our simulation, patients with $\hat{p}_i > 0.5$ are classified to treatment performing subgroup and patients with $\hat{p}_i \leq 0.5$ are classified to treatment non-performing subgroup. The R code for the customized loss function and evaluation function can be found in supplementary materials.

For all scenarios, we generate 1000 replicated datasets. In each replicated dataset we simulate data for 5500 patients, and we use 500 patients as the training set and use the remaining 5000 as the validation set. In each simulated dataset, we compute the proposed value function estimate in (4) using validation set. Note the RMST is computed with time in month. Greater value function value indicates subgroups with greater magnitude of subgroup-treatment interaction effect are identified. In addition, since we know the true subgroup membership for each patient, we calculate the accuracy rate as the percentage of correctly predicted subgroup memberships in the validation set. Sensitivity and specificity are also calculated based on validation set. As a side product of the proposed method and VT, variable importance can be calculated from tree boosting and the classification random forests. We calculate the ranks of $S_1$ and $S_2$ if available in the training set using the default "Gain" index in the xgboost and Gini index for the classification random forests. For Lasso and ROWSi, we order the variables by their shrunken coefficients and find the ranks of $S_1$ and $S_2$ if their coefficients are not shrunk to zero. If their coefficients are shrunk to zero, then we report their ranks as number of non-zero coefficients + median number of zero coefficients. For SGBT, we first compute the relative influence of each covariate under the models for the experimental treatment and control group, respectively. We then order all covariates by their absolute values of the difference in the relative influences between the two models and record the ranks for $S_1$ and $S_2$.

The simulation results under setting 1 are summarized in Table 1 and simulation results under setting 2 are summarized in Table 2. In scenario 1 and scenario 2 when the subgroups are separable by simple linear boundaries, the modified covariate method provides the greatest value function estimates and highest prediction accuracies. The proposed method provides slightly less accuracy and value function estimate compared to SGBT in scenario 1 but outperforms SGBT in scenario 2. The two gradient boosting trees-based methods both outperform the VT and ROWSi by a noticeable amount. When a "U" shaped boundary is presented (scenario 3), the proposed method becomes the winner by providing a value function estimate of 9.78 and a prediction accuracy of 0.97. The SGBT comes the second while the modified covariate method, VT and ROWSi' value function estimates and prediction power drop sharply: VT provides a value function estimate of 2.99 and a prediction accuracy of 0.53, the modified covariate method and ROWSi provide a value function estimate close to 0 and a prediction accuracy of 0.5, suggesting no prediction power. Under scenario 4, the treatment performing subgroup is bounded in a rectangular "enclave". The proposed method provides a value function estimate of 8.84 and a prediction accuracy of 0.87 while all the other methods provide limited prediction power. The SGBT, the method also utilizes the gradient boosting trees but is built upon the proportional hazard model, experiences a significant performance drop with a value function estimate of 4.79 and a prediction accuracy of 0.62. Similarly, the proposed method outperforms the other methods in scenario 5 and 6 by a significant amount in the presence of more irregular boundary.

The observations made from scenario 3-6 suggest that when the subgroup boundaries become more irregular, the performance of the regression based modified covariate method and ROWSi drop rapidly while the proposed method could still provide decent prediction power. The SGBT and VT methods also underperform the proposed method across most of the scenarios examined here. When several prognostic variables contribute to the variation in outcome (setting 2), decline in performance is observed for all methods. Notably, the modified covariate method proposed by Tian et al.[20] is not immune to the presence of prognostic effects even though the method nominally only estimates the interactions effects through modified covariates. Similar in setting 1, the proposed method outperforms others when boundaries become more irregular.

# 4 Real Data Analysis

We apply the proposed method to an AIDS clinical trial data from ACTG175[48] for the illustration purpose. The ACTG 175 is a randomized clinical trial to compare monotherapy with zidovudine (ZDV) or didanosine (DDI) with combination therapy with ZDV and DDI or ZDV and zalcitabine in adults infected with the human immunodeficiency virus type I whose CD4 T cell counts were between 200 and 500 per cubic millimeter. We subset the study population to treatment arms for ZDV + DDI and DDI monotherapy. The subset has 1083 patients with 522 patients in ZDV + DDI arm and 561 patients in DDI monotherapy arm. As in[32], we include 12 covariates in addition to the treatment indicator. The 12 covariates include five continuous variables: age, weight, Karnofsky score, CD4 count at baseline and CD8 count at baseline, and seven binary covariates: hemophilia, homosexual activity, history of intravenous drug use, race, gender, antiretroviral history, and symptomatic status. The goal is to stratify the overall population into subgroups that patients may or may not benefit from ZDV + DDI relative to DDI alone.

To reduce overfitting, we use cross-validation to estimate patients' optimal treatment. Specifically, we partition the data into 5 roughly equal-sized sets based on original order of the observations in the dataset. At each iteration, we pick a different set as the validation set and perform the proposed method to the remaining 4 sets of the data to develop the subgroup membership prediction model. We then predict the subgroup memberships for patients in the validation set and stratify them into ZDV + DDI performing (ZDV+DDI > DDI) subgroup and ZDV + DDI non-performing (ZDV+DDI <= DDI) subgroup. A cutoff c of 0.5 is used. This way patients are not involved in the training process for the prediction model to estimate their subgroup memberships. Among the 1083 patients, 800 patients are classified to the ZDV+DDI performing subgroup, 74% of overall population. In Figure 3, we plot the Kaplan–Meier survival curves for the two treatment arms in the overall population and two subgroups identified by the proposed method: ZDV+DDI > DDI and ZDV+DDI <= DDI. From the plots, in the ZDV+DDI > DDI subgroup the curve for ZDV+DDI arm stays above the DDI arm curve and there is a greater gap between the two curves than it is in the overall population. In the ZDV+DDI <= DDI subgroup, there is not much difference between the two curves and the DDI arm has a slightly better survival than ZDV+DDI arm. We also estimate the value function using (4) based on the subgroups identified by the proposed method and the other methods, based on cross-validation. We conduct cross-validation 1000 times and compute the average and standard deviation of the value function estimates. The results are summarized in Table 3. The subgroups identified by the proposed method has the largest estimated value function, which is a measure of subgroup-treatment interaction effect and is the only one with a positive value meaning ZDV+DDI combination's benefit over DDI monotherapy in the identified treatment performing subgroup is greater than it in the treatment non-performing subgroup. The negative value function estimated from other methods suggest there could be a mismatch between optimizing treatment effect at individual level and maximizing the differential treatment effect at the subgroup level. In contrast, by directly searching for maximized differential treatment effect at subgroup level, the recommended treatment should have the overall benefit over the alternative in both subpopulations identified by the proposed method. In other words, precision medicine should have a sensible population interpretation as well, and that motivates our value function.

Difference in RMST and hazard ratio between two treatment arms across the two subgroups identified by the proposed method based on cross-validation are summarized in Table 4. Finally, we fit the model to the overall sample and plot the Kaplan–Meier survival curves for the two identified subgroups in Figure S1. The variable importance from the tree boosting is summarized in the Figure S2 in supplementary materials. The weight, age and Karnofsky score are the top 3 most important variables to determine the subgroup membership. The above findings suggest ZDV+DDI combination could be preferred to DDI monotherapy for most of the patients but there could be a subgroup of patients who may not benefit from the combination.

## 5 Discussion

Tree-based ensemble methods which enjoy flexible model structure have unique advantages in precision medicine. In this paper we propose a nonparametric method to search for subgroup membership scores by maximizing a value function that directly reflects the subgroup-treatment interaction effect. The simulation results suggest that the proposed method could outperform existing methods in some scenarios when subgroup boundaries become irregular. As an alternative value function, one could follow Qian and Murphy's[23] strategy and sets the value function as in (6) and estimate it as below,

$$\hat{E}\left[\frac{YI(A=\tau(Z))}{P(A|Z)}\right] = \left\{(\sum_{i=1}^n p_i) \int_0^{t^*} \tilde{S}(t, A=1, \boldsymbol{p}) \, dt + [\sum_{i=1}^n (1-p_i)] \int_0^{t^*} \tilde{S}(t, A=0, 1-\boldsymbol{p}) \, dt\right\} \quad (12)$$

We include additional two terms for $E[Y|A \neq \tau]$ in the proposed objective function so that the value function can be used as a measure of subgroup-treatment interaction effect. We compare the value function estimator in (12) to the value function estimator in (4) through simulation and the value function in (4) consistently provides slightly better prediction power and therefore is proposed.

The gradient tree boosting algorithm as applied in the proposed method enables the direct search for the "propensity" of treatment-recommending subgroup membership. The use of population-level gradient for each subject-specific subgroup membership score in the boosting algorithm has not been previously proposed in the personalized medicine literature based on our research. We believe this technique has further potential in a broader class of problems.

The proposed method has limitations. It is primarily designed for a two-armed randomized clinical trial with censored survival data. For observational studies when prognostic variables could correlate with the treatment assignment, the signal captured by the proposed value function may not be due to heterogenous responses to treatments. Extension to multiple treatment setting could be made by modifying the value function to $\hat{V} = \sum_{k=1}^{K} \left[ (\sum_{i=1}^{n} p_{ki}) \int_0^{t^*} S(t, A = k, \boldsymbol{p}_k) \, dt \right]$ for K treatment arms with $\sum_{k=1}^{K} p_{ki} = 1$. Moreover, extension to continuous and binary outcome could be made by replacing RMST with alternative estimator, such as sample mean, for $E(Y|A, \tau)$. The proposed method also suffers from the presence of prognostic variables. A value function that integrated from stratified value functions across prognostic variables defined subgroups could reduce the influence of prognostic effects. In addition, the estimated subgroup membership score is a continuous quantity used to determine subgroup membership with a cutoff c. In simulation study and real data application we use c=0.5 and it leads to decent subgroup identification results. But 0.5 may not be the optimal value and in practice we may be only confident to declare subgroup memberships to patients with scores well greater or less than 0.5.

In practice, researchers often want to make statistical inference about the identified subgroups and associated differential treatment effect. Permutation test might be used for this purpose. For example, one could first optimize and estimate the value function based on observed sample and then repeat randomly "shuffled" datasets a large number of times and at each time generate and save the maximized value function estimate. If the value function estimate based on observed data is greater than the 1-α quantile of the value function estimates from randomly permuted datasets, one could conclude that the identified differential treatment effect is indeed present. These extensions will be the subjects of our future research. Lastly the proposed method is not immune to common problems in statistical learning such as false positive and overfitting and it may not be easy to interpret the subgroups identified as the prediction model could be based on multiple trees. The proposed method should be used as an exploratory tool and it is crucial to have an independent validation set to verify the findings.

DATA AVAILABILITY STATEMENT

The data used in Section 4 is available in the R package "speff2trial" at https://CRAN.R-project.org/package=speff2trial

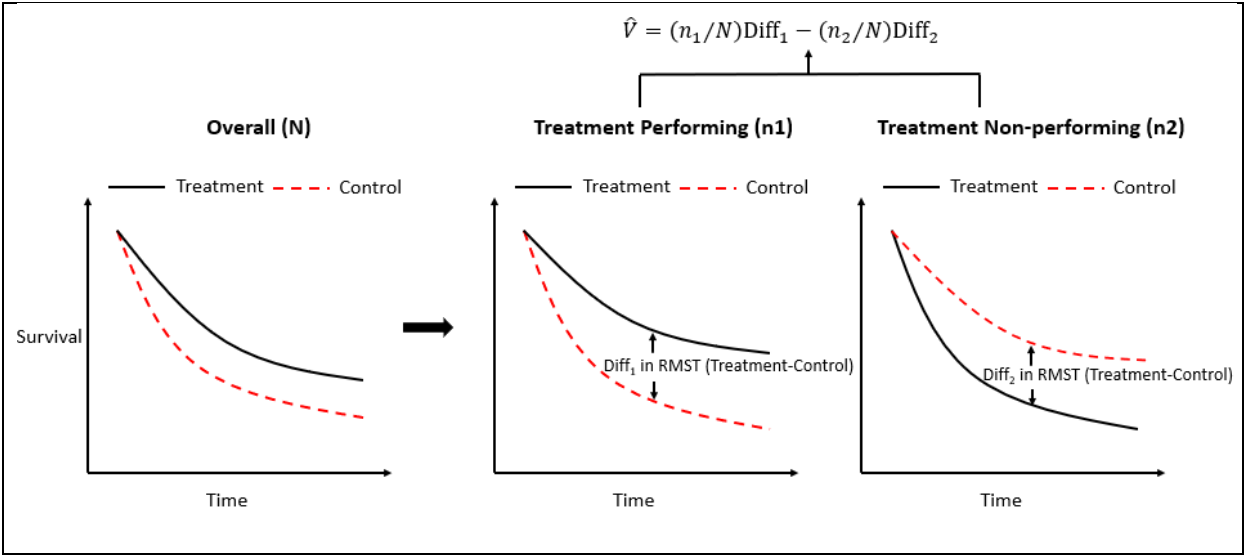

**Figure 1**. Visualization of the value function as a measure of the differential treatment effects across subgroups identified. The value function is a weighted integration of difference in RMST estimates, a measure of gaps between treatment arm survival curve and control arm survival curve, across the two identified subgroups weighted by subgroup prevalence.

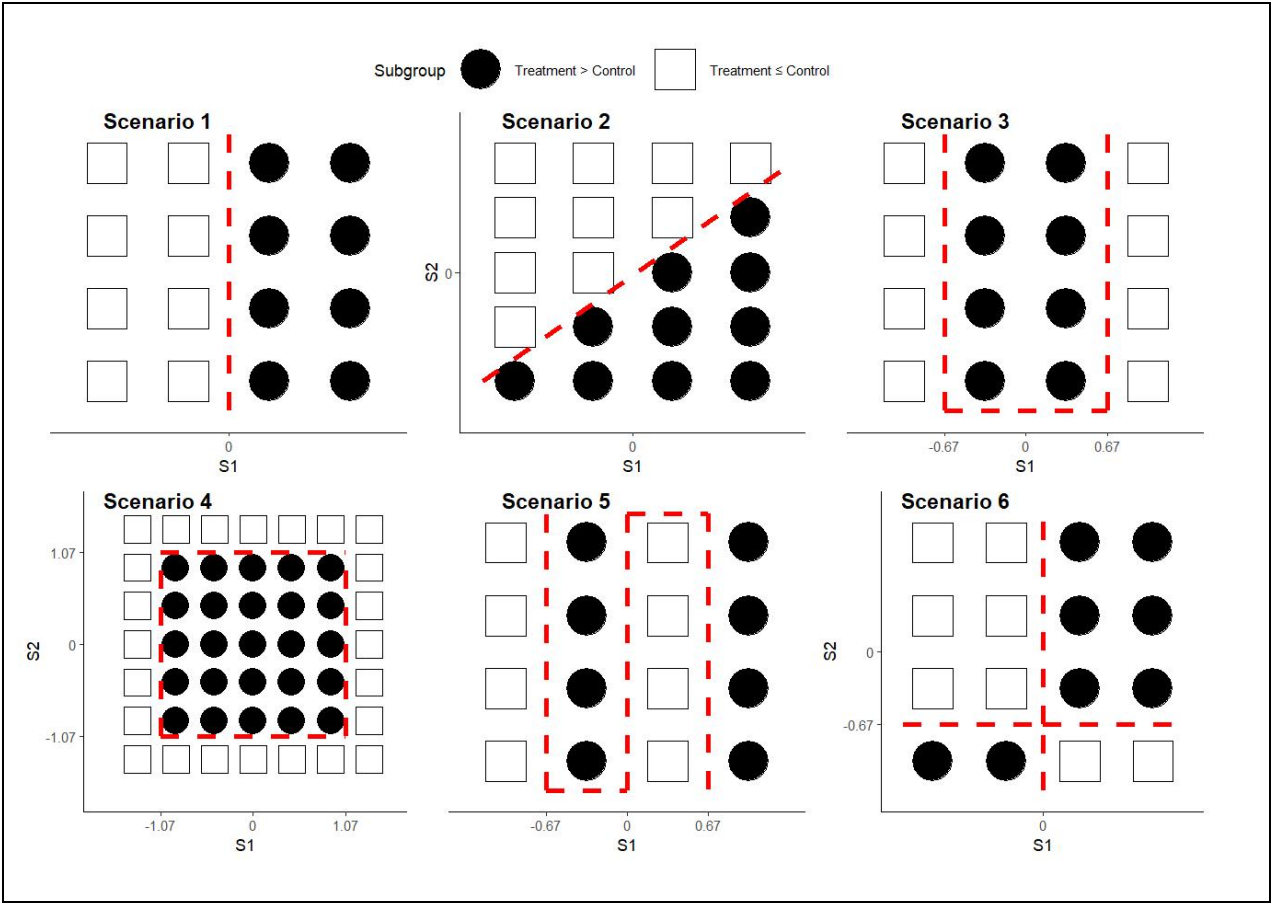

**Figure 2**. Six scenarios of underlying subgroup patterns.

**Table 1** Simulation results when no prognostic factors are presented (setting 1) for value function estimates, classification accuracy, sensitivity, specificity based on validation set, and predictive variable ranking(s) based on training set

| | Method | | | | |
|---|---|---|---|---|---|
| | Proposed Method | SGBT | Modified Covariate | Virtual Twins | ROWSi |
| **Scenario 1** | | | | | |
| $\hat{V}(\tau)$ (s.d.) | 7.19 (0.35) | 7.46 (0.31) | 7.54 (0.26) | 4.45 (1.11) | 6.54 (1.06) |
| Accuracy (s.d.) | 0.93 (0.03) | 0.94 (0.04) | 0.96 (0.03) | 0.67 (0.08) | 0.86 (0.07) |
| Sensitivity (s.d.) | 0.94 (0.05) | 0.98 (0.04) | 0.96 (0.04) | 0.99 (0.02) | 0.78 (0.14) |
| Specificity (s.d.) | 0.93 (0.06) | 0.89 (0.09) | 0.97 (0.03) | 0.36 (0.17) | 0.94 (0.08) |
| Rank ($S_1$, $S_2$) | (1.00, NA) | (1.00, NA) | (1.00, NA) | (1.00, NA) | (1.11, NA) |
| **Scenario 2** | | | | | |
| $\hat{V}(\tau)$ (s.d.) | 6.73 (0.57) | 5.98 (0.99) | 8.16 (0.26) | 1.51 (1.26) | 6.43 (1.42) |
| Accuracy (s.d.) | 0.82 (0.04) | 0.75 (0.07) | 0.95 (0.03) | 0.52 (0.03) | 0.82 (0.07) |
| Sensitivity (s.d.) | 0.86 (0.06) | 1.00 (0.01) | 0.95 (0.03) | 0.99 (0.02) | 0.74 (0.14) |
| Specificity (s.d.) | 0.77 (0.08) | 0.49 (0.14) | 0.95 (0.03) | 0.04 (0.07) | 0.90 (0.12) |
| Rank ($S_1$, $S_2$) | (1.50, 1.50) | (1.50, 1.5) | (1.48, 1.52) | (1.81, 1.58) | (2.13, 2.01) |
| **Scenario 3** | | | | | |
| $\hat{V}(\tau)$ (s.d.) | 9.78 (0.34) | 9.08 (0.66) | 0.80 (1.16) | 2.99 (0.82) | 0.14 (0.90) |
| Accuracy (s.d.) | 0.97 (0.02) | 0.91 (0.04) | 0.50 (0.01) | 0.53 (0.02) | 0.50 (0.01) |
| Sensitivity (s.d.) | 0.99 (0.01) | 1.00 (0.00) | 0.92 (0.22) | 1.00 (0.00) | 0.62 (0.41) |
| Specificity (s.d.) | 0.95 (0.04) | 0.82 (0.09) | 0.08 (0.22) | 0.07 (0.05) | 0.38 (0.40) |
| Rank ($S_1$, $S_2$) | (1.00, NA) | (1.00, NA) | (27.33, NA) | (1.11, NA) | (26.11, NA) |
| **Scenario 4** | | | | | |
| $\hat{V}(\tau)$ (s.d.) | 8.84 (0.44) | 4.79 (0.58) | 1.14 (1.56) | 3.42 (1.19) | 0.35 (1.05) |
| Accuracy (s.d.) | 0.87 (0.02) | 0.62 (0.03) | 0.52 (0.01) | 0.54 (0.01) | 0.51 (0.02) |
| Sensitivity (s.d.) | 0.99 (0.02) | 1.00 (0.00) | 0.97 (0.07) | 1.00 (0.00) | 0.77 (0.32) |
| Specificity (s.d.) | 0.75 (0.04) | 0.20 (0.06) | 0.03 (0.08) | 0.04 (0.03) | 0.23 (0.32) |
| Rank ($S_1$, $S_2$) | (1.66, NA) | (1.62, NA) | (28.25, NA) | (7.54, NA) | (26.84, NA) |
| **Scenario 5** | | | | | |
| $\hat{V}(\tau)$ (s.d.) | 9.74 (1.22) | 1.73 (2.70) | 4.24 (1.33) | 2.83 (1.39) | 2.10 (1.76) |
| Accuracy (s.d.) | 0.94 (0.08) | 0.55 (0.10) | 0.64 (0.05) | 0.53 (0.03) | 0.58 (0.06) |
| Sensitivity (s.d.) | 0.98 (0.02) | 1.00 (0.00) | 0.84 (0.12) | 1.00 (0.00) | 0.70 (0.26) |
| Specificity (s.d.) | 0.91 (0.14) | 0.11 (0.21) | 0.45 (0.10) | 0.05 (0.05) | 0.46 (0.30) |
| Rank ($S_1$, $S_2$) | (1.00, NA) | (1.00, NA) | (1.17, NA) | (1.33, NA) | (9.83, NA) |
| **Scenario 6** | | | | | |
| $\hat{V}(\tau)$ (s.d.) | 9.56 (1.69) | 2.20 (3.29) | 3.53 (2.11) | 0.04 (0.46) | 1.90 (2.03) |
| Accuracy (s.d.) | 0.90 (0.10) | 0.63 (0.09) | 0.59 (0.02) | 0.58 (0.01) | 0.59 (0.04) |
| Sensitivity (s.d.) | 0.98 (0.03) | 1.00 (0.00) | 0.96 (0.04) | 1.00 (0.00) | 0.89 (0.17) |
| Specificity (s.d.) | 0.80 (0.22) | 0.10 (0.21) | 0.07 (0.09) | 0.00 (0.00) | 0.16 (0.23) |
| Rank ($S_1$, $S_2$) | (2.65, 2.20) | (1.92, 1.09) | (2.30, 22.77) | (6.09, 6.03) | (16.87, 26.69) |

**Table 2** Simulation results when prognostic factors are presented (setting 2) for value function estimates, classification accuracy, sensitivity, specificity based on validation set, and predictive variable ranking(s) based on training set

| | Method | | | | |
|---|---|---|---|---|---|
| | Proposed Method | SGBT | Modified Covariate | Virtual Twins | ROWSi |
| **Scenario 1** | | | | | |
| $\hat{V}(\tau)$ (s.d.) | 3.24 (0.78) | 3.72 (0.41) | 3.76 (0.96) | 0.84 (0.98) | 1.57 (1.57) |
| Accuracy (s.d.) | 0.79 (0.08) | 0.84 (0.05) | 0.86 (0.12) | 0.56 (0.10) | 0.65 (0.14) |
| Sensitivity (s.d.) | 0.79 (0.13) | 0.93 (0.08) | 0.84 (0.18) | 0.92 (0.15) | 0.48 (0.35) |
| Specificity (s.d.) | 0.78 (0.14) | 0.74 (0.11) | 0.88 (0.12) | 0.20 (0.25) | 0.82 (0.30) |
| Rank ($S_1$, $S_2$) (s.d.) | (1.23, NA) | (1.00, NA) | (1.87, NA) | (8.29, NA) | (10.08, NA) |
| **Scenario 2** | | | | | |
| $\hat{V}(\tau)$ (s.d.) | 4.00 (0.82) | 3.85 (0.51) | 5.81 (0.48) | 1.15 (1.03) | 3.39 (1.92) |
| Accuracy (s.d.) | 0.71 (0.05) | 0.67 (0.03) | 0.86 (0.06) | 0.53 (0.04) | 0.68 (0.12) |
| Sensitivity (s.d.) | 0.72 (0.11) | 0.85 (0.07) | 0.86 (0.06) | 0.85 (0.18) | 0.62 (0.26) |
| Specificity (s.d.) | 0.70 (0.12) | 0.49 (0.10) | 0.86 (0.07) | 0.21 (0.22) | 0.75 (0.26) |
| Rank ($S_1$, $S_2$) | (4.16, 2.31) | (1.00, 5.1) | (1.79, 1.34) | (15.33, 20.63) | (9.23, 7.51) |
| **Scenario 3** | | | | | |
| $\hat{V}(\tau)$ (s.d.) | 6.28 (1.06) | 5.46 (0.78) | 0.99 (0.82) | 1.51 (1.08) | 0.38 (0.70) |
| Accuracy (s.d.) | 0.89 (0.08) | 0.82 (0.04) | 0.51 (0.01) | 0.55 (0.05) | 0.50 (0.01) |
| Sensitivity (s.d.) | 0.93 (0.10) | 0.94 (0.06) | 0.41 (0.26) | 0.91 (0.13) | 0.51 (0.41) |
| Specificity (s.d.) | 0.85 (0.11) | 0.69 (0.10) | 0.61 (0.25) | 0.19 (0.18) | 0.49 (0.40) |
| Rank ($S_1$, $S_2$) | (1.10, NA) | (1.00, NA) | (24.17, NA) | (8.69, NA) | (26.15, NA) |
| **Scenario 4** | | | | | |
| $\hat{V}(\tau)$ (s.d.) | 5.17 (0.62) | 1.40 (1.66) | 1.17 (1.39) | 2.93 (1.02) | 0.69 (1.16) |
| Accuracy (s.d.) | 0.81 (0.05) | 0.53 (0.01) | 0.52 (0.01) | 0.54 (0.01) | 0.52 (0.01) |
| Sensitivity (s.d.) | 0.97 (0.05) | 1.00 (0.00) | 0.92 (0.18) | 1.00 (0.00) | 0.95 (0.12) |
| Specificity (s.d.) | 0.64 (0.09) | 0.02 (0.03) | 0.09 (0.18) | 0.04 (0.02) | 0.06 (0.13) |
| Rank ($S_1$, $S_2$) | (2.57, 2.88) | (1.84, 1.44) | (28.11, 28.07) | (12.49, 13.05) | (26.72, 26.92) |
| **Scenario 5** | | | | | |
| $\hat{V}(\tau)$ (s.d.) | 2.80 (0.56) | 0.92 (1.06) | 1.31 (0.84) | 1.06 (0.94) | 0.99 (0.95) |
| Accuracy (s.d.) | 0.74 (0.09) | 0.52 (0.03) | 0.60 (0.06) | 0.51 (0.02) | 0.55 (0.05) |
| Sensitivity (s.d.) | 0.91 (0.06) | 1.00 (0.00) | 0.75 (0.22) | 1.00 (0.02) | 0.94 (0.11) |
| Specificity (s.d.) | 0.57 (0.18) | 0.04 (0.06) | 0.45 (0.20) | 0.02 (0.03) | 0.15 (0.18) |
| Rank ($S_1$, $S_2$) | (1.18, NA) | (1.00, NA) | (4.26, NA) | (5.16, NA) | (12.67, NA) |
| **Scenario 6** | | | | | |
| $\hat{V}(\tau)$ (s.d.) | 2.71 (0.93) | 0.18 (0.63) | 1.72 (0.81) | 0.49 (0.94) | 0.86 (1.02) |
| Accuracy (s.d.) | 0.71 (0.15) | 0.58 (0.01) | 0.60 (0.04) | 0.58 (0.01) | 0.59 (0.02) |
| Sensitivity (s.d.) | 0.90 (0.07) | 1.00 (0.00) | 0.87 (0.14) | 1.00 (0.01) | 0.95 (0.08) |
| Specificity (s.d.) | 0.44 (0.32) | 0.01 (0.02) | 0.23 (0.22) | 0.01 (0.01) | 0.08 (0.13) |
| Rank ($S_1$, $S_2$) | (10.65, 11.99) | (1.67, 4.85) | (5.38, 26.57) | (11.21, 16.62) | (15.67, 26.71) |

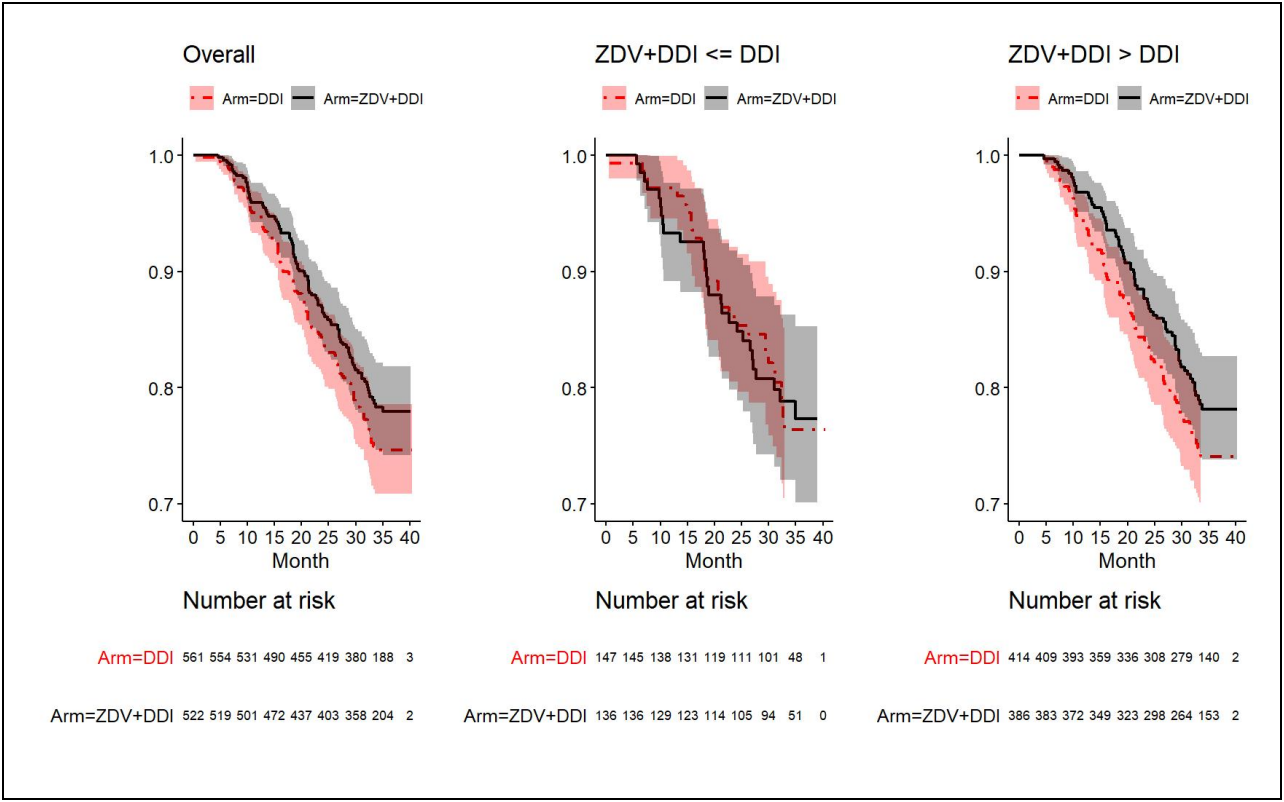

**Figure 3.** The Kaplan–Meier survival curves in the overall population and subgroups identified by the proposed method using cross-validation for ACTG175. **Left plot**: Overall population. **Middle plot**: subgroup identified to be not benefited from ZDV+DDI compared to DDI monotherapy. **Right plot**: subgroup identified to be benefited from ZDV+DDI compared to DDI monotherapy.

Table 3 Summary of differential treatment effect across subgroups identified by different methods based on 1000 times randomly repeated cross-validation for ACTG175. RMST estimates are based on time in month

|  | Differential Treatment effect between Subgroups | | | | |
| --- | --- | --- | --- | --- | --- |
|  | Proposed Method | SGBT | Modified Covariate | VT | ROSWi |
| $\hat{V}(\tau)$ (s.d.) | 0.60 (0.47) | -0.01 (0.52) | -0.64 (0.46) | -0.17 (0.43) | -0.44 (0.46) |

**Table 4** Summary of restricted mean survival time (RMST) difference and hazard ratio between two treatment arms within the subgroups identified by the proposed method using cross-validation for ACTG175

| Population | Difference in RMST (Month) (95% CI) | Hazard Ratio (95% CI) |
|---|---|---|
| Overall | 0.85 (-0.28, 1.99) | 0.84 (0.65, 1.09) |
| ZDV+DDI>DDI | 1.25 (-0.07, 2.56) | 0.79 (0.59, 1.07) |
| ZDV+DDI<=DDI | -0.28 (-2.39, 1.83) | 0.98 (0.59, 1.65) |